# A single crystal high-temperature pyrochlore antiferromagnet

J. W. Krizan[1*] and R. J. Cava[1]

[1]Department of Chemistry, Princeton University, Princeton, NJ 08544, USA

* Corresponding Author: jkrizan@princeton.edu

**Abstract**

We report the magnetic characterization of the frustrated transition metal pyrochlore $NaCaCo_2F_7$. This material has high spin $Co^{2+}$ in $CoF_6$ octahedra in a pyrochlore lattice, and disordered non-magnetic Na and Ca on the large-atom sites in the structure. Large crystals grown by the floating zone method were studied. The magnetic susceptibility is isotropic, the Co moment is larger than the spin-only value, and in spite of the large Curie Weiss theta (-140 K), freezing of the spin system, as characterized by peaks in the *ac* and *dc* susceptibility and specific heat, does not occur until around 2.4 K. This yields a frustration index of $f = -\theta_{CW}/T_f \approx 56$, an indication that the system is highly frustrated. The observed entropy loss at the freezing transition is low, indicating that magnetic entropy remains present in the system at 0.6 K. The compound may be the realization of a frustrated pyrochlore antiferromagnet with weak bond disorder. The high magnetic interaction strength, strong frustration, and the availability of large single crystals makes $NaCaCo_2F_7$ an interesting alternative to rare earth oxide pyrochlores for the study of geometric magnetic frustration in pyrochlore lattices.

**Introduction**

The pyrochlore lattice is one of the canonical magnetically frustrated lattices [1]. The $Ln_2Ti_2O_7$ (Ln = rare earth ion) pyrochlores in particular have been extensively studied due to the availability of large single crystals and the wide range of magnetic ground states they display, which arise from the comparable strengths of their magnetic interactions and crystal field effects; long range ordered, spin ice, spin glass, and spin liquid states are observed at low temperatures. [2,3] The A and B ions in the $A_2B_2X_7$ pyrochlores are each on their own sublattice of corner sharing tetrahedra, and thus systems with magnetic ions on either the pyrochlore A or the B sites are of interest as potentially frustrated materials. With virtually all of the focus on oxides, relatively few fluoride pyrochlores have been discovered and fewer yet have been magnetically characterized. Here we report the characterization of single crystals of a fully fluorine-based $A_2B_2F_7$ pyrochlore, $NaCaCo_2F_7$ [4], with a high effective moment for the Co ($P_{eff} \approx 6\ \mu_b$) a high Curie Weiss theta (-140 K), and no spin freezing until 2-3 K, indicating that the material is highly magnetically frustrated. The magnetic data and the disorder of the nonmagnetic off-site A ions Na and Ca imply that $NaCaCo_2F_7$ may be an embodiment of a pyrochlore antiferromagnet with weak bond disorder which adds a perturbation to the Co-Co exchange interactions and has been described theoretically [5–7]. This has been proposed to explain the enigmatic properties of $Y_2Mo_2O_7$ [8]. $NaCaCo_2F_7$ can be grown as large single crystals by the floating zone method, and promises the existence of a large new family of high quality magnetic transition metal single crystal materials for expansion of the field of study of geometric magnetic frustration on pyrochlore lattices.

**Experimental**

Single crystals of $NaCaCo_2F_7$ were prepared in an optical floating zone furnace, and full structural characterization of the material was performed by single crystal X-ray diffraction at 300 K (See Supplemental Material at [*URL will be inserted by publisher*]). Temperature and field dependent *ac* and *dc* magnetization measurements of the oriented crystals were made using the ACMS option in a Quantum Design Physical Property Measurement System (PPMS) and Quantum Design superconducting quantum interference device (SQUID) equipped Magnetometer (MPMS-XL-5). Heat capacity measurements were made using the heat relaxation method in the PPMS; 4-6 mg single crystal samples were mounted on a nonmagnetic sapphire stage with Apiezon N grease. A single crystal of the non-magnetic analogue $NaCaZn_2F_7$ was used for the subtraction in order to estimate the magnetic contribution to the specific heat in $NaCaCo_2F_7$.

**Results**

The results of the single crystal structural determination of $NaCaCo_2F_7$ are presented in the Supplemental Material at [*URL will be inserted by publisher*], Tables I and II. The material is a stoichiometric $A_2B_2F_7$ pyrochlore, where Na and Ca are completely disordered on the A-site and $Co^{2+}$ fully occupies the B site. There is no indication of any long range or short range Na-Ca ordering (see the reciprocal lattice plane image in Figure 1), nor is any A-site ion/B-site ion mixing detected. The *x*-parameter of one fluorine atom in the structure dictates the geometry of the F octahedron around the $Co^{2+}$ ion and thus the crystal field. *x*=0.3125 would indicate a perfect octahedron; for $NaCaCo_2F_7$, *x*=0.333, indicating that the octahedra are distorted in shape, even though the average Co-F distances are all equivalent (Figure 1). The octahedra are compressed along a <111> type direction with two different F-Co-F bond angles, 97.7° and 82.3°, rather than the ideal 90.0° value. Adjacent Co octahedra that connect to form a frustrated Co-tetrahedron all have the axis of compression pointing towards the center of the Co-tetrahedron. This level of distortion is in line with that observed in the oxide pyrochlores. [3] The powder diffraction pattern for $NaCaCo_2F_7$ is shown in Figure 1; a single crystal of $NaCaCo_2F_7$ cut from the larger floating zone single crystal boule is also shown in the Figure.

Figure 2 shows the temperature dependent magnetic susceptibility (χ is defined as M/H where H = 2000 Oe; M vs. H is linear in this field range, see below) of $NaCaCo_2F_7$ with the field applied parallel to the [111] crystallographic direction (lower right inset). The Curie-Weiss fit (main panel) is to the 150-300 K data and is of the type χ=C/(T-θ) where C is the Curie constant and θ the Weiss temperature. The upper left inset shows the magnetization of $NaCaCo_2F_7$ as a function of field up to 9 T at 2 K for the [100], [110], and [111] directions; the dependence is highly linear for all directions and not nearly saturated at 9 T. The temperature dependent data shows that $NaCaCo_2F_7$ has dominantly antiferromagnetic interactions, is highly frustrated, and has a very large effective moment per Co. $NaCaCo_2F_7$ presents a Weiss temperature of approximately -140 K and an effective moment per cobalt of approximately 6.1 $\mu_b$, which is much larger than the spin only moment (3.87 $\mu_b$) for S=3/2 $Co^{2+}$ and closer to what would be expected for the presence of a full orbital contribution (6.63 $\mu_b$).

Figure 3 further characterizes the magnetism of $NaCaCo_2F_7$ at temperatures near the magnetic freezing, which happens at approximately 2.4 K in spite of the much higher Curie Weiss temperature of -140 K. The top panel shows the temperature dependent *ac* susceptibility from 2-4 K at a number of different frequencies. (The data for the [111] direction is presented for comparison to the behavior in the spin ice $Dy_2Ti_2O_7$ [9,10].) No features are seen in the *ac* susceptibility between 5 and 300 K (data not shown). A peak is seen in the real phase of the *ac* susceptibility in the 2-4 K temperature range, which is verified to reflect the spin freezing, as seen in the specific heat measurements described below. A small 0.16 K shift of the peak in the *ac* susceptibility is seen as a function of frequency, indicative of glassiness in the spin freezing transition. The shift of the peak temperature as a function of frequency can be described by the expression $\frac{\Delta T_f}{T_f \Delta log\omega}$, which is used to characterize spin-glass and spin-glass-like materials [11–13]. The value obtained for $NaCaCo_2F_7$ is 0.029, which is slightly higher than expected for a canonical spin glass and in the range of what is expected for an insulating spin glass [11,12]; it is an order of magnitude lower than that expected for a super paramagnet and an order of magnitude greater than that of cluster-glasses [11,13]. To further characterize the glassiness of the transition, the Volger-Fulcher law can be applied. This law takes into account that interactions between clusters of spins complicate a spin glass transition and make it more than just a simple thermally activated process. The Volger-Fulcher law gives a relation between the freezing temperature ($T_f$) and the frequency (*f*): $T_f = T_0 - \frac{E_a}{k_b} \frac{1}{\ln(\tau_0 f)}$; the intrinsic relaxation time ($\tau_0$), the activation energy of the process ($E_a$), and, "the ideal glass temperature" ($T_0$) [11,13] can be extracted. Due to having only five data points and limited temperature resolution (the total observed temperature shift of the peak is only 0.16 K), no physical result could be obtained by fitting all parameters simultaneously, however, good fits could be achieved by fixing $\tau_0$ to a physically realistic value. The intrinsic relaxation time ($\tau_0$) usually falls between $1x10^{-13}$ (conventional spin glasses) and $1x10^{-7}$ (super-paramagnets, cluster glasses). As such, fits with $\tau_0$ = $1x10^{-13}$, $1x10^{-12}$ and $1x10^{-11}$ respectively gave $E_a$ = 1.3 $x10^{-3}$, 1.0 $x10^{-3}$, and 7.9 $x10^{-4}$ eV and $T_0$ = 2.10, 2.17, and 2.24 K. The fit where $\tau_0$ = $1x10^{-12}$, a mid-range value of the same order as that found ($\tau_0 = 2.5 \times 10^{-12}$) in the insulating spin glass $Fe_2TiO_5$ is presented in the bottom panel of Figure 3. [14] The "ideal glass temperature" can be interpreted as either relating to the cluster interaction strength in a spin glass, or relating to the critical temperature of the underlying phase transition that is dynamically manifesting at $T_f$ [11]. The middle panel of Figure 3 shows the zero field cooled (ZFC) and field cooled (FC) *dc* susceptibility in a field of 200 Oe applied parallel to the [110] direction of $NaCaCo_2F_7$ in the range of 2-3.5 K. Similar behavior is seen for the [100] and [111] directions (not shown). The small difference in the zero field cooled vs. field-cooled susceptibilities are as is expected for a magnetic freezing transition that has glassy character. This bifurcation at approximately 2.4 K is very close to the $T_0$ indicated by the Volger-Fulcher fits and is suppressed with field; it is not detectable with an applied field of 500 Oe or greater. We take the maximum in the *dc* susceptibility at 2.4 K as an estimate of the spin freezing temperature ($T_f$). This data can be used to parameterize the behavior of the material at the spin freezing point; the frustration index [1] $f = -\theta_{CW}/T_f$ can be determined as ~ 56, indicating that $NaCaCo_2F_7$ is highly magnetically frustrated.

Further characterization of the freezing of the frustrated spins can be seen in the heat capacity of $NaCaCo_2F_7$ and its comparison to that of the non-magnetic analogue $NaCaZn_2F_7$ shown in the upper panel and inset of Figure 4. No scaling of the $NaCaZn_2F_7$ data was employed. The low temperature comparison of the two data sets is shown in the middle panel and inset, which shows a much higher heat capacity for the Co compound vs. the Zn compound, which is a reflection of the entropy loss from the freezing of the magnetic system. The magnetic heat capacity given in these panels is obtained from the subtraction of the data for Zn from the data for Co. A broad transition can be observed in the low temperature magnetic heat capacity at approximately the same temperature that is seen for the transition observed in the *ac* susceptibility. The inset to the bottom panel gives a close-up view of the magnetic heat capacity below the transition. It is sometimes possible to learn about the elementary excitations and infer the ground state by looking at the heat capacity below the transition [1]. Curvature is evident in the current data plotted as $C_{mag}/T$ vs. T, and it is also not linear on $T^2$ or $T^{3/2}$ scales, thus we cannot come to any conclusion about the low temperature spin excitations from the current data. The integration of the magnetic heat capacity over temperature for the data that is not affected by the subtle differences in the phonon contributions of the Zn and Co analogs that occur above 75 K, allows for an estimate of the temperature dependence of the magnetic entropy loss. This is given in the bottom panel. The magnetic entropy appears to saturate by 50 K, and is compared to the expected value for S=3/2 Heisenberg and Ising systems, Rln(2S+1), and Rln(2), respectively. A significant amount of magnetic entropy is frozen out at low temperatures in the main transition, but neither Rln(2S+1) nor Rln(2) entropy is released on heating, implying that there is considerable residual magnetic entropy in the $NaCaCo_2F_7$ system at 0.6 K. The proximity of the integrated entropy loss to Rln(2) suggests that the magnetic system may be Ising-like, but further work will be required to make that determination conclusively.

**Conclusion**

We have observed no long range magnetic ordering at low temperatures in $NaCaCo_2F_7$ in spite of the large $Co^{2+}$ moment and high antiferromagnetic interaction strengths inferred from the Curie Weiss fits to the magnetic susceptibility data; the magnetic heat capacity shows only a broad maximum between 2 and 3 K and what appears to be substantial residual entropy at 0.6 K. Raman and infrared scattering studies of the analogous NaCaMg$_2$F$_7$ pyrochlore showed that the Ca-Na disorder on the pyrochlore A-site relaxed the vibrational selection rules around the pyrochlore B site, showing that the local B-site electrostatics were affected by the off-site A-site disorder [15]. The same type of off-site disorder exists in NaCaCo$_2$F$_7$, which therefore may lead to random magnetic bonds in the B-site to B-site interactions, and in turn to a spin glass ground state. Theoretical studies of the Heisenberg antiferromagnet on the pyrochlore lattice indicate that weak randomness in the exchange interactions (i.e. magnetic bond disorder) can precipitate a spin glass ground state, with the spin glass temperature set by the strength of the bond disorder [5–7]; whether this is the case in NaCaCo$_2$F$_7$ would be of interest for future experimental and theoretical study. Finally, we observe that fluoride-based pyrochlore and Kagome systems [16] are promising new avenues for research in frustrated magnets. These materials accommodate new ions, in particular the transition elements with their typically high magnetic moments and strong magnetic interactions, in classical frustrating geometries. The availability of large, high quality single crystals of the fluoride pyrochlores should facilitate future studies.

**Acknowledgements**

The authors thank C. Broholm, O. Tchernyshyov, K. Ross and R. Moessner for helpful discussions. This research was conducted under the auspices of the Institute for Quantum Matter at Johns Hopkins University, and supported by the U. S. Department of Energy, Division of Basic Energy Sciences, Grant DE-FG02-08ER46544.

Figure Captions:

Figure 1. (Color on line) Upper left panel: a (110) slice of a NaCaCo$_2$F$_7$ single crystal cut from a larger floating zone boule. Top center panel: a precession image from single crystal diffraction of the (0kl) plane, which illustrates the lack of any additional long range structural ordering. Top right panel: the CoF$_6$ octahedra. The compression of the octahedra is along a <111> type axis. While the average Co-F bond lengths are identical, there are two different F-Co-F bond angles as indicated by the color of the hypotenuse of the triangle; purple indicates a bond angle of 82.3° and green an angle of 97.7°. Bottom panel: a representative powder pattern of NaCaCo$_2$F$_7$ confirms that the bulk material of the crystal boule is an A$_2$B$_2$F$_7$ pyrochlore.

Figure 2. (Color on line) The *dc* susceptibility of NaCaCo$_2$F$_7$, with the applied field parallel to the [100], [110], and [111] crystallographic directions. Main panel: the inverse susceptibility of the [111] direction, the Curie-Weiss fit, and the fitted parameters. Lower right panel: the raw susceptibility of the [111] direction. Upper left panel: magnetization as a function of applied field at 2 K for all three crystallographic directions.

Figure 3. (Color on line) Top panel: the temperature dependence of the real part of the *ac* susceptibility in an applied field of 20 Oe parallel to the [111] direction as a function of frequency. The behavior is parameterized in the fit to the Volger-Fulcher law in the bottom panel. Middle panel: bifurcation of the field cooled and zero field cooled *dc* susceptibility with an applied field of 200 Oe applied parallel to the [110] direction.

Figure 4. (Color on line) Upper panel: the raw heat capacity for single crystals of NaCaCo$_2$F$_7$ and the nonmagnetic analogue NaCaZn$_2$F$_7$. Upper panel inset: the magnetic heat capacity of NaCaCo$_2$F$_7$ as determined by the subtraction of the heat capacity of the Zn analogue. Middle panel and inset: a close-up of the low temperature data given in the upper panel. Lower panel inset: a close-up of the magnetic heat capacity below the transition. Lower panel: the integration of the magnetic heat capacity yields an entropy that saturates below 50 K to a value substantially less than the Ising limit of R ln(2) or the Heisenberg limit of R ln(2S+1).

Figure 1:

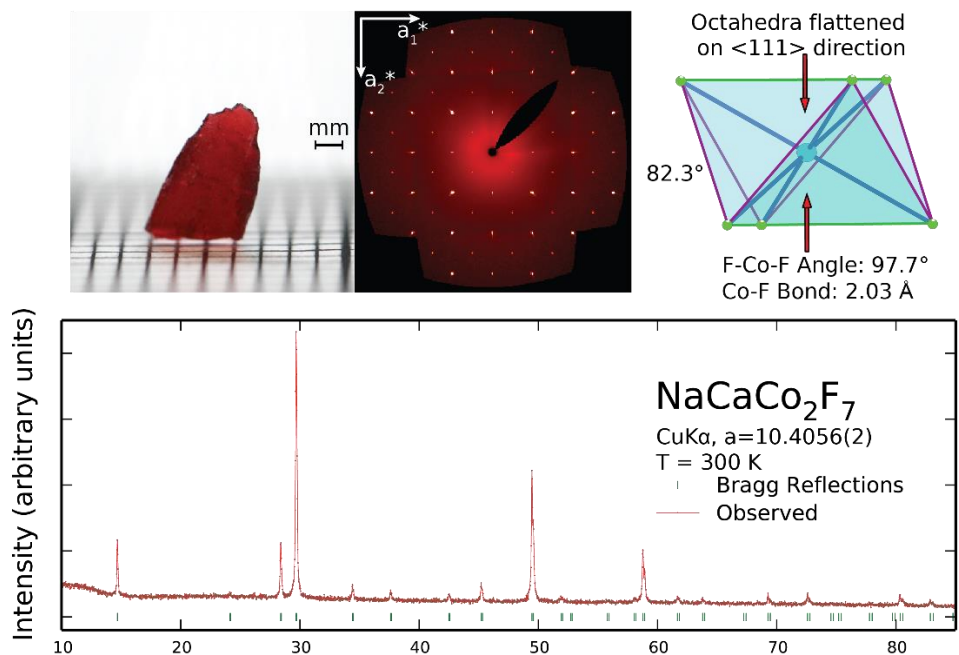

Figure 2:

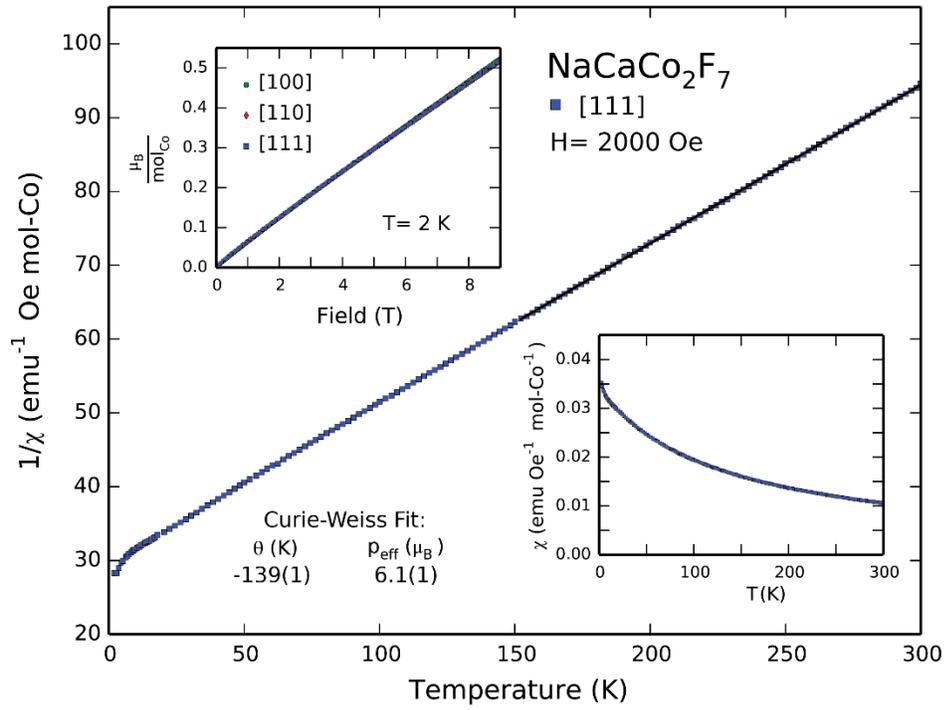

Figure 3:

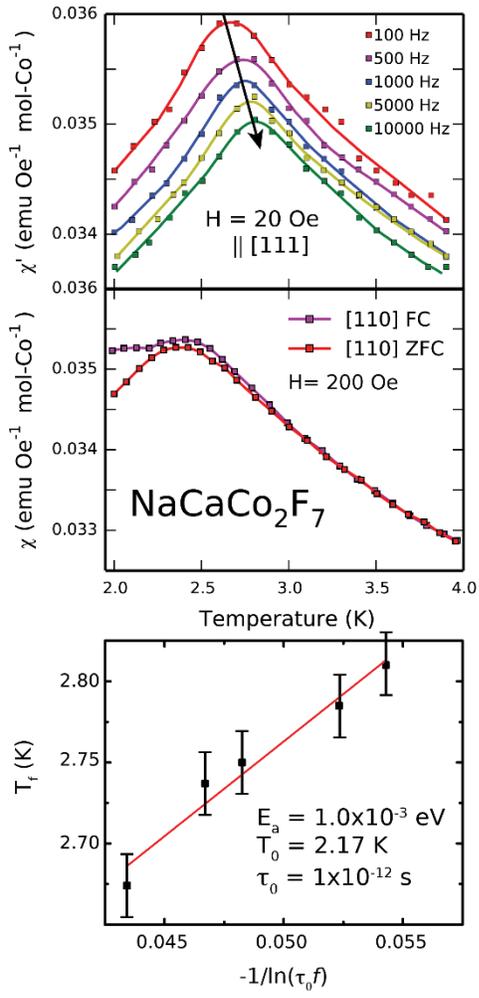

Figure 4:

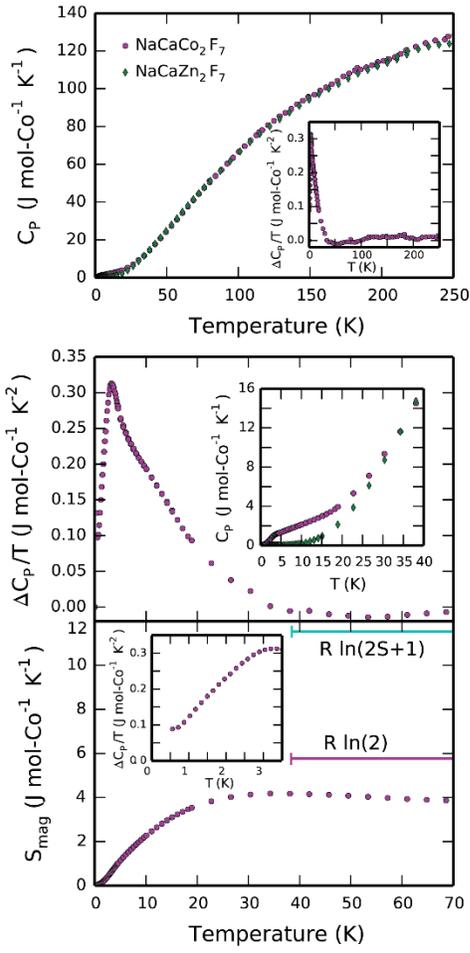

**Supplemental Material**

**Crystal Growth**

The feed rod for the crystal growth process was prepared from a stoichiometric mixture of the binary elemental fluorides. After pressing the unreacted fluorides in a hydrostatic press at 70 MPa to form a 3/16" diameter feed rod, drying was done under a dynamic vacuum for 8 hours. The feed rod was then transferred directly to the floating zone furnace (Crystal Systems Corporation FZ-T-10000-H-VI-VPO, equipped with 300 W halogen bulbs) where the crystals were grown under 8.5 bar of Ar pressure. Growth at rates between 3 mm/hr. and 5 mm/hr. were successful. Both $NaCaCo_2F_7$ and $NaCaZn_2F_7$ (prepared as a non-magnetic analog to $NaCaCo_2F_7$) were prepared by the same method. Single crystal discs of $NaCaCo_2F_7$ of several mm dimension cut from the larger boules were oriented using a Laue camera (Multiwire Laboratories, MWL110), such that magnetic fields could be applied parallel to the [100], [110], and [111] crystal directions. All crystals employed for the magnetic characterization were cut from the same floating zone boule and immediately next to each other.

**Structure determination**

Single crystal X-ray diffraction (SXRD) determination of the crystal structure of $NaCaCo_2F_7$ was performed on a Bruker diffractometer equipped with an Apex II detector and graphite monochromated Mo Kα radiation. Data collection utilized the Bruker APEXII software package and subsequent reduction and cell refinement were performed using Bruker SAINT [S1]. The crystal structure was determined through the use of SHELXL-2013 as implemented through the WinGX software suite [S2,S3]. PXRD at 300 K, using a Bruker D8 Focus with diffracted beam graphite monochromator (Cu Kα), confirmed the bulk material in the boules of both $NaCaCo_2F_7$ and $NaCaZn_2F_7$ to have the pyrochlore structure. PXRD patterns were fit with the Thompson-Cox-Hastings pseudo Voight profile convoluted with axial divergence asymmetry through the FullProf software suite [S4]. For the fitting of $NaCaZn_2F_7$, the structural model was the same as used for $NaCaCo_2F_7$ though with the substitution of Zn in the structural model. Unit cells of 10.4056 (2) Å and 10.3773 (2) Å were found for bulk pieces of the $NaCaCo_2F_7$ and $NaCaZn_2F_7$ floating zone crystal boules by least squares fitting of the PXRD patterns, in good agreement with the single crystal diffraction and previous reports [S5,S6].

**Table 1: Single Crystal Data and Structural Refinement for NaCaCo$_2$F$_7$**

| | |
|---|---|
| Formula weight | 313.923 g/mol |
| Crystal System | Cubic |
| Space Group | $F\,d\,\bar{3}\,m$ (227, origin 2) |
| Unit Cell | a=10.4189(9) Å |
| Volume | 1131.0(3) Å$^3$ |
| Z | 8 |
| Radiation | Mo Kα |
| T | 293 K |
| Absorption Coefficient | 6.922 |
| F(000) | 1184 |
| Reflections collected/unique | 3041/87 R$_{int}$ = 0.0275 |
| Data/Parameters | 87/11 |
| Goodness-of-fit | 1.277 |
| Final R indices [I>2σ(I)] | R$_1$=0.0118 , wR$_2$=0.0331 |
| Largest diff. peak and hole | 0.19 and -0.20 e A$^{-3}$ |

**Table 2: Atomic positions and anisotropic thermal displacement parameters for NaCaCo$_2$F$_7$**

| | Site | x | y | z | Occ. | U11 | U22 | U33 | U23 | U13 | U12 |
|---|---|---|---|---|---|---|---|---|---|---|---|
| Na | 16d | 0.5 | 0.5 | 0.5 | 0.5 | 0.0178(3) | 0.0178(3) | 0.0178(3) | -0.00374(19) | -0.00374(19) | -0.00374(19) |
| Ca | 16d | 0.5 | 0.5 | 0.5 | 0.5 | 0.0178(3) | 0.0178(3) | 0.0178(3) | -0.00374(19) | -0.00374(19) | -0.00374(19) |
| Co | 16c | 0 | 0 | 0 | 1 | 0.0081(3) | 0.0081(3) | 0.0081(3) | -0.00036(8) | -0.00036(8) | -0.00036(8) |
| F(1) | 8b | 0.375 | 0.375 | 0.375 | 1 | 0.0134(6) | 0.0134(6) | 0.0134(6) | 0 | 0 | 0 |
| F(2) | 48f | 0.33289(14) | 0.125 | 0.125 | 1 | 0.0231(8) | 0.0172(5) | 0.0172(5) | 0.0081(5) | 0 | 0 |